\documentclass[%
 reprint,
superscriptaddress,
 amsmath,amssymb,
 aps,
prb,
]{revtex4-2}
\usepackage{graphicx}
\usepackage{dcolumn}
\usepackage{bm}
\usepackage[unicode=true,
 bookmarks=false,
 breaklinks=true,pdfborder={0 0 1},backref=false,colorlinks=true]
 {hyperref}
\hypersetup{
 pdfcreator={},pdfproducer={LaTeX with hyperref},linkcolor=blue,anchorcolor=blue,citecolor=blue,filecolor=red,menucolor=red,urlcolor=blue,pdfstartview=FitV,pdfhighlight=/I,pdfpagelayout=TwoColumns,hypertexnames=true}

\begin{document}
\title{Dynamic paramagnon-polarons in altermagnets}
\author{Charles R. W. Steward}
\affiliation{Institute for Theory of Condensed Matter, Karlsruhe Institute of Technology,
76131 Karlsruhe, Germany}
\author{Rafael M. Fernandes}
\affiliation{School of Physics and Astronomy, University of Minnesota, Minneapolis,
Minnesota 55455, USA}
\author{J\"org Schmalian}
\affiliation{Institute for Theory of Condensed Matter, Karlsruhe Institute of Technology,
76131 Karlsruhe, Germany}
\affiliation{Institute for Quantum Materials and Technologies, Karlsruhe Institute
of Technology, 76126 Karlsruhe, Germany}
\begin{abstract} 
The combined rotational and time-reversal symmetry breakings that define an altermagnet lead to an unusual $d$-wave (or $g$-wave) magnetization order parameter, which in turn can be modeled in terms of multipolar magnetic moments.
Here, we show that such an altermagnetic order parameter couples to the dynamics of the lattice even in the absence of an external magnetic field.
This coupling is analogous to the non-dissipative Hall viscosity and  describes the stress generated  by a time-varying strain under broken time-reversal symmetry.
We demonstrate that this effect generates a hybridized paramagnon-polaron mode, which allows one to assess altermagnetic excitations directly from the phonon spectrum. Using a scaling analysis, we also demonstrate that the dynamic strain coupling strongly affects the altermagnetic phase boundary, but in different ways in the thermal and quantum regimes. In the ground state for both 2D and 3D systems, we find that a hardening of the altermagnon mode leads to an extended altermagnetic ordered regime, whereas for non-zero temperatures in 2D, the softening of the phonon modes leads to increased fluctuations that lower the altermagnetic transition temperature. In 3D even at finite temperatures the dominant effect is the suppression of quantum fluctuations. We also discuss the application of these results to standard ferromagnetic systems.
\end{abstract}
\maketitle
\section{Introduction}
\label{sec:introduction}
Ferromagnets and antiferromagnets are states of broken time-reversal symmetry with finite uniform or staggered magnetic dipole moments arising from uniform or periodic configurations of the electronic spin. A rather different type of magnetic order, which has recently received significant attention \cite{Smejkal2020crystal,mazin2021prediction,Smejkal2022b,Turek2022,Urru2022,bhowal2022magnetic,mazin2023altermagnetism,feng2022anomalous,voleti2020multipolar,mosca2022modeling,betancourt2023spontaneous,winkler2023theory,yuan2021prediction,liu2022spin,bai2023efficient,vsmejkal2022anomalous,yang2021symmetry,jiang2023enumeration}, is multipolar magnetism, which exhibits more complex patterns despite having zero net (staggered) magnetization. Time-reversal symmetry breaking in these cases is due to the formation of magnetic quadrupoles, octupoles, toroidal moments, or higher-order configurations of dipole moments whose averaged magnetization vanishes by symmetry. When the symmetry characterizing the ordered state involves a combination of rotations and time reversal, the system is known as an altermagnet \cite{Smejkal2020crystal,mazin2021prediction,Smejkal2022,Smejkal2022b,Turek2022,Urru2022,bhowal2022magnetic,mazin2023altermagnetism}. In many of the cases studied so far, the altermagnetic order parameter corresponds to a $d$-wave or $g$-wave magnetization, which in turn can also be expressed in terms of multipolar magnetic moments \cite{Kusunose2018,hayami2020bottom,yang2021symmetry}. More broadly, these altermagnetic states correspond to even-parity spin-triplet Pomeranchuk instabilities of the Fermi liquid \cite{Pomeranchuk1958}. In particular, the nematic-spin-nematic state proposed and investigated in Refs. \cite{Oganesyan2001,Wu2007} has an order parameter corresponding to a $d$-wave magnetization, which is the same order parameter proposed for the various candidate altermagnetic compounds \cite{Smejkal2022,Smejkal2022b,bhowal2022magnetic}.

In dipolar magnetic materials, such as standard ferro- and antiferromagnets, changes in the lattice parameter modify the exchange interaction between the spins. Such a magneto-elastic coupling has important consequences, e.g. the emergence of hybrid magnon-acoustic phonon modes in the magnetically ordered state, called magnon-polarons \cite{flebus2017magnon}. In multipolar magnetic materials, however, different types of coupling between the magnetization $M_i$ and the strain $\varepsilon_{ij}$ are allowed \cite{aoyama2023piezomagnetic,caciuffo2011multipolar,rimmler2023atomic,bhowal2022magnetic,kikkawa2016magnon}. 
Formally, while the magneto-elastic effect is associated with the magnetostriction response tensor $N_{ijkl}$, defined as $\varepsilon_{ij} = N_{ijkl} M_k M_l$, several higher-order multipolar magnetic states, including altermagnets, have a non-zero piezomagnetic response tensor $\Lambda_{ijk}$, defined by $\varepsilon_{ij} = \Lambda_{ijk} M_k$ \cite{aoyama2023piezomagnetic}. 
A direct consequence of the piezomagnetism of these altermagnets is that application of a magnetic field should lead to a (possibly symmetry-breaking) lattice distortion in the ordered state -- or, alternatively, a lattice distortion should induce a non-zero magnetization \cite{Patri2019,Fisher2021}. 

In this regard, multipolar order plays a role  that is in some aspects analogous to nematic order \cite{fang2008theory,Fernandes2014,borzi2007formation}, in particular when the multipolar order is associated with a rotational symmetry of the lattice, e.g. octupolar magnetic order. The difference is of course that nematic order does not break time-reversal symmetry which, as we will see, gives rise to fundamental differences.

In this paper, we show that a subset of altermagnets (and even some standard ferromagnets) displays another non-trivial coupling between magnetic and elastic degrees of freedom. By this effect, a strain mode $\varepsilon_{\Gamma^+}$ that transforms as the $\Gamma^+$ irreducible representation of the relevant point group couples to the momentum operator $\pi$ that is canonically conjugate to the fluctuating multipolar order parameter $\phi$ that characterizes the altermagnetic state:
\begin{equation}
   \mathcal{H}^{\rm dyn}_{\rm c}=\frac{\lambda_0}{2}c^2\int d^{3}
   \boldsymbol{x} \varepsilon_{\Gamma^+}(\boldsymbol{x})\pi(\boldsymbol{x}).
   \label{eq:dyncoupl_intro}
\end{equation}
This effect is reminiscent of the Hall viscosity response \cite{Volovik1984,Avron1995,Hughes2011,Bradlyn2012,Rao2020}, 
which describes the stress $\sigma_{ij}$ generated (non-dissipatively) by a time-varying strain, $\sigma_{ij} = \eta_{ijkl} \frac{\partial\varepsilon_{kl}}{\partial t}$. In our case, it is the multipolar magnetic moment $\phi$ characterizing the altermagnetic state that is generated by the time-changing strain. 

Importantly, this dynamic paramagnon-phonon coupling is fundamentally different from the more standard static coupling generated by magnetostriction, which is given by \cite{flebus2017magnon,Brataas2019}

\begin{equation}
   \mathcal{H}^{\rm mag-el}_{\rm c} = \int d^{3}
   \boldsymbol{x} N_{ijkl} \varepsilon_{ij}(\boldsymbol{x})\phi_k(\boldsymbol{x})\phi_l(\boldsymbol{x}) ,
\label{eq:mag_el}
\end{equation}where $N_{ijkl}$ is the magnetoelastic tensor introduced above and $\phi_i$ are the components of the (staggered) magnetization. Such a static coupling also hybridizes the magnetic and elastic collective excitations, giving rise to magnon-polaron modes \cite{flebus2017magnon,Brataas2019}. This effect is most relevant when the phonon and magnon branches cross, where it opens a gap and it can even promote non-trivial topological effects \cite{Go2019,Kwon2020}. In contrast, as we show in this paper, the dynamic coupling of Eq. (\ref{eq:dyncoupl_intro}) plays an important role even if the paramagnon and phonon dispersions do not cross. We note that while the dynamic coupling is not allowed in an antiferromagnet due to momentum conservation, it may arise in ferromagnets with appropriate underlying crystalline symmetries. As we will show later, this is the case for orthorhombic ferromagnets. In contrast, the magnetoelastic coupling of Eq. (\ref{eq:mag_el}) has at least one term allowed regardless of the underlying crystalline symmetry, $N_{iijj}$, corresponding to a volume change caused by magnetic order.

\begin{figure*}
         \includegraphics[width=0.7\textwidth]{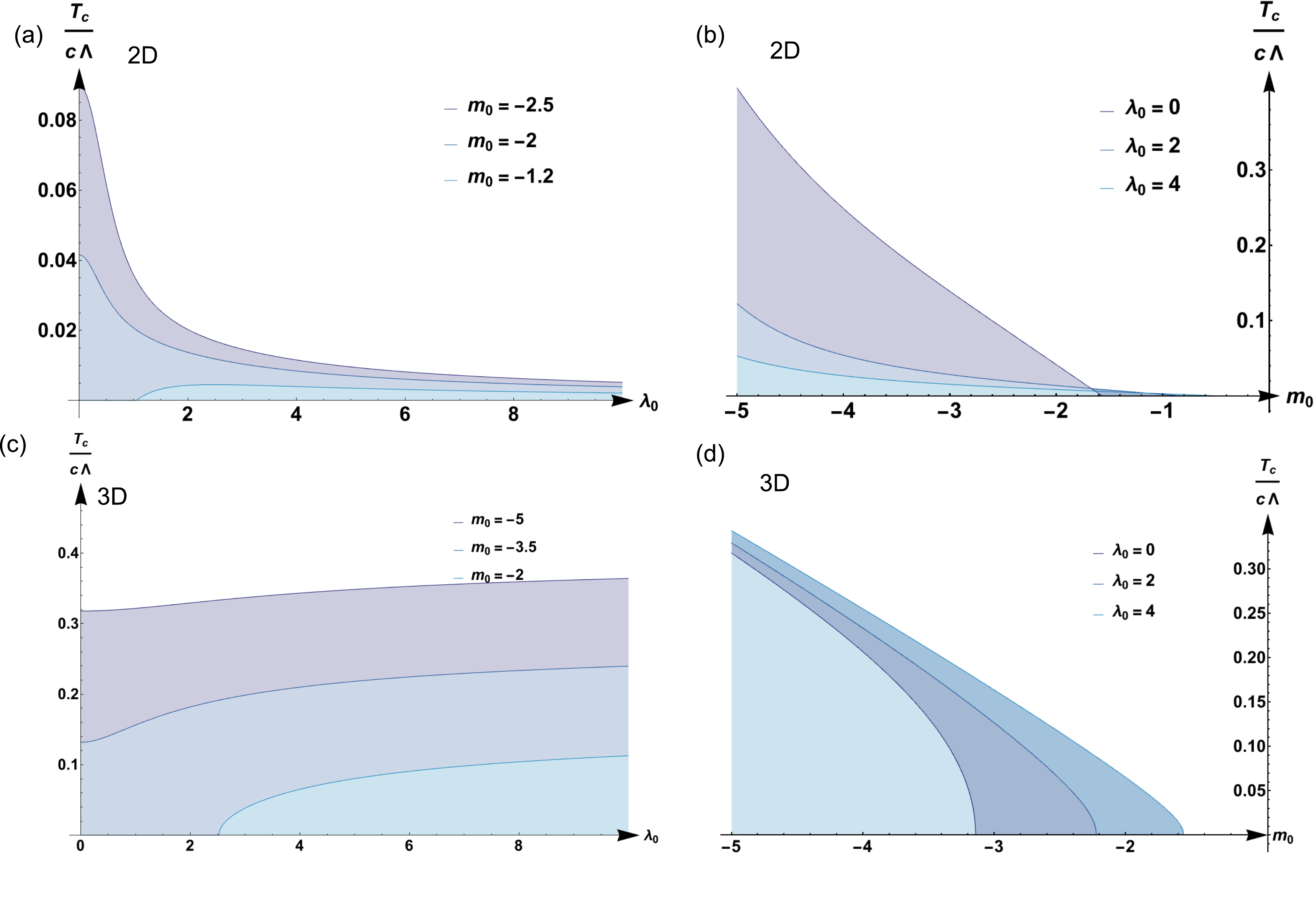}
         \label{fig::Phase1}
        \caption{(a) Altermagnetic transition temperature $T_c$ in a 2D system as a function of the dynamic strain coupling constant $\lambda_0$ of Eq.~\eqref{eq:dyncoupl_intro} for different values of the bare altermagnon mass $m_0=r_0/\Lambda^2$. Note that $m_0 < 0$ ($m_0 > 0$) defines the ordered (disordered) state without the coupling to the lattice and the $\phi^4$ mode coupling. The dynamic coupling to phonons suppresses order at finite temperature, but enhances or even induces order at zero temperature. (b) Altermagnetic $T_c$ in a 2D system as a function of $m_0$ for different values of $\lambda_0$. (c) and (d) show the same information as panels (a) and (b) respectively but for a 3D system.
        These plots refer to the regime $c>v_{L,T}$; the opposite regime is shown in Fig.\ref{fig::Phasev}. }
        \label{fig::Phase}
\end{figure*}

Here, we first study the impact of this dynamic strain-multipolar moment coupling on the elastic-magnetic collective modes of the paramagnetic phase, i.e. before any multipolar magnetic long-range order sets in. We demonstrate the emergence of a hybridized paramagnon-polaron mode, which opens the possibility of detecting the dispersion of the paramagnetic altermagnons (or magnons) directly from the phonon spectrum. Moreover, we also point out that this coupling can be understood as a two-mode squeezing \cite{hong2007one,bogdanovic1979canonical} of the elastic and magnetic multipolar modes.

Second, we investigate how the altermagnetic-to-paramagnetic transition is affected by the dynamic strain coupling. We consider both thermal and quantum fluctuations -- indeed, by tuning appropriate parameters, it is in principle possible to reach a quantum critical point \cite{sachdev1999quantum,vojta2003quantum,stishov2004quantum,schattner2016ising} where the altermagnetic transition temperature vanishes, or a quantum disordered regime where altermagnetic fluctuations affect the ground state without long-range order. Surprisingly, we find distinct effects of the dynamic coupling on thermal and quantum fluctuations. In the important regime where the bare magnon velocity is  larger than the phonon velocities, thermal fluctuations are boosted by the altermagnon-phonon coupling. In 2D these fluctuations are strong enough to suppress the ordering temperature. On the other hand, quantum fluctuations are suppressed by the same coupling, leading to an increased regime of stable altermagnetic ground states. In 3D where thermal fluctuations are weaker, this suppression of quantum fluctuations is the dominant effect and coupling to phonons causes an increase in the ordering temperature. These behaviors are illustrated by the phase diagrams in Fig. \ref{fig::Phase}.

This paper is organized as follows: in Section \ref{sec:dynstraincoupl}, we define our multipolar magnetic order parameter and its coupling to strain. In Section \ref{sec::fieldtheory} we construct a $\phi^4$ theory for the altermagnetic degrees of freedom and write down the elastic theory for the crystal in question. Having done so, we are then able to calculate an effective field theory for the altermagnons, and derive the altermagnon and phonon dispersions and spectral functions, this is presented in Section \ref{sec::Spectral}. In Section \ref{sec::phase} we then perform an RG (renormalization group) calculation for the altermagnon propagator in the crossover regime, yielding the phase diagram for the system. Section \ref{sec:conclusions} contains the summary and conclusions, including the possible extension of our results to certain ferromagnets.

\section{Dynamic coupling between strain and multipolar order}
\label{sec:dynstraincoupl}
We start by defining the multipolar moment of an altermagnet. Consider a magnetic order parameter $\phi$ transforming under an irreducible
representation $\Gamma^-$ which, by construction, is odd under time
reversal, as indicated here by the superscript ``$-$''. Let $\Gamma^-_{J_{\alpha}}$ be the representation of the
magnetic field component $H_{\alpha}$, i.e. the representation according
to which magnetic dipoles transform under the point group operations. If $\Gamma^-=\Gamma^-_{J_{\alpha}}$
we have the usual ferromagnetic order parameter
\begin{equation}
\phi^{\alpha}\left(\boldsymbol{x}\right)\sim\sum_{ab}c_{a}^{\dagger}\left(\boldsymbol{x}\right)J_{ab}^{\alpha}c_{b}\left(\boldsymbol{x}\right).\label{eq:dipole}
\end{equation}
where $J^{\alpha}$ is an angular momentum operator and $a$ and $b$ stand for spin and orbital indices. $c_{a}^{\dagger}\left(\boldsymbol{x}\right)$ and $c_{a}\left(\boldsymbol{x}\right)$ are corresponding electron annihilation and creation operators. In the simplest
case $J^{\alpha}$ is one of the Pauli matrices; in general it also includes
an orbital moment. 
Other representations that are odd under time reversal but different than $\Gamma^-_{J_{\alpha}}$ form higher order multipolar magnetic  order
parameters that behave like (see also Ref. \cite{Kusunose2018})
\begin{equation}
\phi\left(\boldsymbol{x}\right)\sim\sum_{ab}\int d^{3}\boldsymbol{x}'f\left(\boldsymbol{x}'\right)c_{a}^{\dagger}\left(\boldsymbol{x}+\frac{\boldsymbol{x}'}{2}\right)J_{ab}^{\mu}c_{b}\left(\boldsymbol{x}-\frac{\boldsymbol{x}'}{2}\right),\label{eq:general}
\end{equation}
with some form factor $f\left(\boldsymbol{x}\right)$. Clearly, $f\left(\boldsymbol{x}\right) = \delta\left(\boldsymbol{x}\right)$ recovers the ferromagnetic order parameter in Eq. (\ref{eq:dipole}). The same is true if 
$f\left(\boldsymbol{x}\right)$ transforms trivially under point-group operations.
However, other form factors that transform non-trivialy  give rise to higher-order multipoles \cite{hayami2020bottom}. In particular, when $f\left(\boldsymbol{x}\right)$ corresponds to $d$-wave, $g$-wave or $i$-wave form factors, these order parameters describe an altermagnet \cite{Smejkal2022,Smejkal2022b}. The underlying
distribution of the spin (or orbital moment) density in the unit cell is most naturally
a consequence of multiple atoms per unit cell, even in situations where there is only
one electronic band crossing the Fermi surface. In what follows we
focus on these types of multipolar order parameters.

In order to motivate the coupling of strain and multipolar magnetic order, we briefly summarize the established case of strain coupling of a nematic order parameter $\eta$. Suppose $\eta$ transforms under a representation $\Gamma^+$, which is, by definition of a nematic state, time-reversal even. Then it couples to strain 
in the Hamiltonian $\mathcal{H}$ via the nemato-elastic coupling
\begin{equation}
    \mathcal{H}^{\rm n.e.}_{\rm c}=\lambda_{\rm n.e.}\int d^{3}\boldsymbol{x}\varepsilon_{\Gamma^+}(\boldsymbol{x})\eta(\boldsymbol{x}).
    \label{eq:nemelast}
\end{equation}
$\varepsilon_{\Gamma^+}$ is the combination of strain tensor elements that transforms like $\Gamma^+$; see below for examples. This coupling gives rise to a structural distortion at a nematic transition. Even without nematic long-range order one can relate the nematic susceptibility and the elastic constants \cite{fernandes2010effects}. The latter gets softened whenever there is a large nematic susceptibility and a sizeable coupling constant $\lambda_{\rm n.e.}$.

A coupling of the type Eq.~\eqref{eq:nemelast} is not allowed for multipolar order $\phi$, even if it transforms like the elastic strain tensor under the symmetry operations of the crystal. The issue is that strain is even under time reversal and will not directly couple to magnetism. One way to resolve this is by adding an external magnetic field $H_\alpha$. Then a symmetry-allowed coupling of the kind 
\begin{equation}
    \mathcal{H}^H_{\rm c}=\sum_{i}\sum_{\alpha=x,y,z}\lambda^H_{\alpha, i}H_{\alpha}\int d^{3}\boldsymbol{x}\varepsilon_{\Gamma^+_{i}}(\boldsymbol{x})\phi(\boldsymbol{x})
    \label{eq:fieldcoupling}
\end{equation}
emerges, provided $\Gamma^-\in\Gamma^-_{J_{\alpha}}\otimes\Gamma^+_{i}$, i.e. the product representation of  field and strain contains $\Gamma^-$. This is just another way of expressing the piezomagnetic response of a multipolar magnet, $\varepsilon_{ij} = \Lambda_{ijk} M_k$, which makes  explicit the proportionality between the relevant piezomagnetic tensor elements $\Lambda_{ijk}$ and the multipolar magnetic order parameter $\phi$. Couplings of the type Eq.~\eqref{eq:nemelast} and \eqref{eq:fieldcoupling} are static and thus occur for generic, non-dynamic order parameter configurations \cite{Patri2019,aoyama2023piezomagnetic}. 

If one, however, considers the dynamics of the order parameter, one can identify another direct coupling to strain that does not require a finite magnetic field. First, note that an order parameter $\phi(\boldsymbol{x})$ has under rather generic dynamics a conjugated momentum $\pi(\boldsymbol{x})$. In the quantum regime, this implies the canonical commutation relations $\left[\phi(\boldsymbol{x}),\pi(\boldsymbol{x}')\right]=i\hbar  \delta(\boldsymbol{x}-\boldsymbol{x}')$, while in the classical regime the dynamics follows from the corresponding Poisson brackets.  Since  $\phi$ is odd under time reversal, $\pi$ is even and hence transforms as $\Gamma^+$. This implies that strain and multipolar order couple like the one given in Eq.~\eqref{eq:dyncoupl_intro}, provided that there are combinations of the strain tensor $\varepsilon_{ij}$ that transform as $\Gamma^+$. Expressed in terms of an action, this coupling takes, after eliminating the conjugated momentum $\pi$  in favor of the time derivative of the order parameter $\partial_\tau \phi$, the form

\begin{equation}
   \mathcal{S}^{\rm dyn}_{\rm c}=\frac{\lambda_0}{2}\int_0^\beta d\tau \int d^{3}\boldsymbol{x}\varepsilon_{\Gamma^+}(\boldsymbol{x},\tau)\partial_\tau\phi(\boldsymbol{x},\tau).
   \label{eq:dyncoupl}
\end{equation}
 Here, $\lambda_0$ is a coupling constant and $\beta=1/T$ the inverse temperature and $\tau$ the imaginary time.  $c$ given in  Eq.~\eqref{eq:dyncoupl_intro} is a velocity, which will be properly defined below. As we will see below, $\lambda_0$ is dimensionless. Hence, for systems with strong coupling to the lattice, a natural value of the coupling constant is  $\lambda_0\approx 1$.   
 
 In contrast to the piezomagnetic coupling of Eq. (\ref{eq:fieldcoupling}), the interaction Eq.~\eqref{eq:dyncoupl_intro} or, equivalently, Eq.~\eqref{eq:dyncoupl}   does not affect static field configurations. However, it strongly  mixes the dynamics of lattice and magnetic degrees of freedom and  is present even in the absence of an external magnetic field. As we will see, it opens up the possibility to observe dynamical multipolar magnetic fluctuations via Raman \cite{lai2021detection,sugai1989phonon,mironova2019magnon,abdalian1980raman,kamba2014strong} or neutron scattering \cite{chatterji2013phonon,fong1995phonon,pintschovius2005electron,shaw1971investigation,kulda1994inelastic}, even in the magnetically disordered state.  We further motivate such a coupling in Appendix \ref{sec::coupling}. The analysis of the coupling in Eq. \eqref{eq:dyncoupl} is the content of the rest of this paper.

 The formulation in Eq.~\eqref{eq:dyncoupl} allows us to make the aforementioned connection to the Hall viscosity explicit. The relationship between dynamic strain and stress is given by \cite{landau_elasticity}
 \begin{equation}
 \sigma_{ij}=C_{ijkl}\varepsilon_{kl}-\eta_{ijkl}\partial_t \varepsilon_{kl},
 \label{eom_stress}
 \end{equation}
with the usual elastic constants $C_{ijkl}$ and
the viscosity tensor $\eta_{ijkl}$. The second term  takes into account that deformations performed
at a finite speed are dissipative and produce heat.
Indeed, elements of the viscosity tensor that are symmetric under the
exchange of $ij \longleftrightarrow kl$ contribute
to the entropy production. However,  antisymmetric contributions are non-dissipative.
Due to the Onsager reciprocity relation such antisymmetric components
occur as a consequence of broken time-reversal symmetry. Let us consider a system without altermagnetic fluctuations but in an external magnetic field. In the
presence of a finite magnetic field it is allowed for the
antisymmetric components to be non-zero. An example, relevant to the point group $D_{4h}$, which we discuss in detail below, is the Hall viscosity 
\begin{equation}
\eta^{H}\equiv\eta_{xyxx}\left(B_{z}\right)=-\eta_{xxxy}\left(-B_{z}\right)=\cdots.
\end{equation}
The Hall viscosity contribution in the action that yields  the equation of motion Eq. \eqref{eom_stress} is then
\begin{eqnarray}
{\cal S}_{{\rm Hall}} & = & -\frac{1}{2}\eta^{H}\int d\tau d^{3}\boldsymbol{x}\left[\left(\epsilon_{xx}-\epsilon_{yy}\right)2\partial_{\tau}\epsilon_{xy}\right. \nonumber\\
 & - & \left.\left(\varepsilon_{xy}+\varepsilon_{yx}\right)\left(\partial_{\tau}\varepsilon_{xx}-\partial_{\tau}\varepsilon_{yy}\right)\right].
\end{eqnarray}
Comparing this with Eq.~\eqref{eq:dyncoupl} shows that  $\eta^{H}\left(B_{z}\right)\partial_{\tau}\varepsilon_{xy}$  plays the same role as   $ \partial_{\tau}\phi$ if we consider $\Gamma^{+}=B_{1g}$. Since $\eta^{H}\left(B_{z}\right)=-\eta^{H}\left(-B_{z}\right)$  both are odd under time reversal and both transform the same way under
point group operations. The fluctuating field due to the altermagnetic order parameter suffices, which is why
Eq.~\eqref{eq:dyncoupl} does not require an external magnetic field. The dynamics of the order parameter induces non-dissipative stress, in  analogy to the Hall viscosity response.

To proceed, we consider specific crystalline point groups and altermagnetic order parameters. Here for simplicity we first focus on layered but three-dimensional systems.
Let us consider  a tetragonal system with point group $D_{4h}$ \cite{shapiro2015symmetry}. Magnetic order that preserves lattice translations should transform under one of the five irreducible representations of the point group that are odd under time reversal (see also Ref. \cite{Fernandes2023}). Of those, $A_{2g}^{-}$ and $E_{g}^{-}$ correspond to ferromagnetic states with magnetization along the $z$-axis and in the $x-y$ plane, respectively. Those form magnetic dipoles, i.e. usual magnetic order. In addition, one can form higher order moments that transform like $A_{1g}^{-}$, $B_{1g}^{-}$ or $B_{2g}^{-}$. 
Along the $k_z=0$ plane, the form factors of Eq.~\eqref{eq:general} in momentum space are $f_{B_{1g}^{-}}\left(\boldsymbol{k}\right)=\sin k_{x}\sin k_{y}$,
$f_{B_{2g}^{-}}\left(\boldsymbol{k}\right)=\cos k_{x}-\cos k_{y}$,
and $f_{A_{1g}^{-}}\left(\boldsymbol{k}\right)=f_{B_{1g}^{-}}\left(\boldsymbol{k}\right)f_{B_{2g}^{-}}\left(\boldsymbol{k}\right)$. In all cases, the spins point out of the plane.
Note that the $A_{1g}^{-}$ state corresponds to a magnetic dotriacontapole, while the other
two form magnetic octupoles. They can be understood as charge
multipoles, characterized by the form factor $f\left(\boldsymbol{x}\right)$,
which modulate a pseudo-vector $s^{\mu}$ that by itself transforms line
a magnetic dipole \cite{hayami2020bottom}. Thus, if $f\left(\boldsymbol{x}\right)$
describes a quadrupolar (hexadecapolar) distribution, i.e. with angular moment $l=2$ ($l=4$),
then $\phi\left(\boldsymbol{x}\right)$ corresponds to an octupolar (dotriacontapole)
magnetic moment with $j=l+1=3$ ($j=l+1=5$).
To clarify our notation, we use the subscript $f_{\Gamma^-}\left(\boldsymbol{k}\right)$ as the irreducible representation of the order parameter, not of the form factor function $f\left(\boldsymbol{k}\right)$ itself.  From Eq.~\eqref{eq:general} follows  $\Gamma^-= \Gamma_{J_\alpha}^-\otimes \Gamma_f$, where $\Gamma_f$ is the representation of the form factor. The above results follow with $\Gamma_{J_z}=A_{2g}^-$.  

 All three states of multipolar order in $D_{4h}$ are single component states and can be described by an Ising order parameter $\phi$ \cite{dziarmaga2005dynamics}. To be specific, we assume below that $\phi$ transforms according to $B_{1g}^{-}$, which is the altermagnetic order parameter proposed for MnF$_2$ \cite{Smejkal2022,bhowal2022magnetic}; the modifications for the other symmetries are straightforward. 
 The coupling to strain given in Eq.~\eqref{eq:dyncoupl} is then given as 
 \begin{equation}
   H^{\rm dyn}_{\rm c}=\frac{\lambda_0}{2}c^2 \int d^3\boldsymbol{x} \varepsilon_{B_{1g}}(\boldsymbol{x})\pi(\boldsymbol{x}),
    \label{eq:altmelast}
\end{equation}
 where $\varepsilon_{B_{1g}}(\boldsymbol{x})=\varepsilon_{xx}(\boldsymbol{x})-\varepsilon_{yy}(\boldsymbol{x})$.
 
To give another example, consider the octahedral group $O_{h}$.
Dipolar magnetic order in this group transforms as the three-dimensional
irreducible representation $T_{1g}^{-}$, amounting to ferromagnetic order
along the crystalline axes. In addition, there are four multipolar
order parameters that do not break inversion symmetry: $A_{1g}^{-}$, $A_{2g}^{-}$, $E_{g}^{-}$, and $T_{2g}^{-}$. Strain transforms either as  $A_{1g}^{+}$, amounting to volume changes $\epsilon_{xx}+\epsilon_{yy}+\epsilon_{zz}$;
$E_{g}^{+}$, with doublet $\left( 2\epsilon_{zz}-\epsilon_{xx}-\epsilon_{yy}, \epsilon_{xx}-\epsilon_{yy} \right)$;
and $T_{2g}^{+}$ with triplet $\left( \epsilon_{xy},\epsilon_{xz}, \epsilon_{yz}\right)$.
Thus, if the order parameter transforms like $A_{2g}^{-}$, there exists
no strain field that can dynamically couple via the conjugated momentum, while a
 field induced coupling to $T_{2g}^{+}$-strain via Eq.~\eqref{eq:fieldcoupling}  is allowed. 
Both types of couplings are allowed for the other two order parameter options. 

The group $O_h$ is relevant for Pr$M_{2}$Al$_{20}$ systems
with transition metal $M$ where multipolar order has been discussed extensively \cite{sakai2011kondo,sakai2012superconductivity,matsubayashi2012pressure,freyer2018two}. However, the candidate state for multipolar magnetic order is $A_{2g}^-$ (ferro-octupolar order) and no strain combination transforms as this irreducible representation.
Hence, the coupling,  Eq.~\eqref{eq:dyncoupl_intro} is not realized here.

\section{Manifestations of the dynamic strain coupling on the collective modes}
\label{sec:qcp}

The dynamic coupling in Eq.~\eqref{eq:dyncoupl} between strain and a multipolar magnetic order parameter is particularly interesting in the regime where quantum fluctuations are strong, i.e. in the quantum critical regime where the ordering temperature has been suppressed to zero. In this regime we can describe the multipolar order in terms of a long-wavelength collective field theory. We will first explore the implications of the interaction Eq. \eqref{eq:altmelast} in the case of a single-component order parameter with point group $D_{4h}$. In the next subsection we will briefly summarize the theory of multipolar order and fluctuations without the dynamic coupling to strain. Then we will add the strain coupling and analyze  the resulting  coupled problem within a renormalized Gaussian  approach. While the latter is justified by the fact that we operate at the upper critical dimension, we go beyond the Gaussian theory and include critical fluctuations using a one-loop renormalization group approach in the next section. This method is particularly suitable to determine the impact of the dynamic elastic coupling on the phase boundary of multipolar magnetism.

\subsection{Field theory for coupled multipolar and elastic degrees of freedom}\label{sec::fieldtheory}

We briefly summarize the collective field theory of multipolar order in the absence of coupling to elastic degrees of freedom.
We consider an insulating system and analyze the regime near its quantum critical point, i.e. the regime where the quantum dynamics of the order parameter is most important. The single-component system is described in terms of an Ising order parameter and governed by the action
\begin{equation}
    \mathcal{S}_\phi=\frac{1}{2}\int_x \phi(x)\left(r_0-c^{-2}\partial_\tau^2-\nabla^2\right)\phi(x)+u\int_x \phi(x)^4.
    \label{eq:Sphi}
\end{equation}
Here $x=(\boldsymbol{x},\tau)$ combines the spatial coordinates and the imaginary time, while $\int_x\cdots=\int d^3\boldsymbol{x} d\tau \cdots $. The parameter $r_0$, which is the mass term for the altermagnons, tunes the system through the quantum critical point. $c$ is the altermagnon velocity, which is of order of the typical magnetic interaction $J$ times the lattice constant. The coefficient $u$ penalizes large-amplitude fluctuations and bounds the action. 

Before proceeding, we note that the situation is slightly different for two-component multipolar order parameters, like the $E_g^-$ state of $O_h$ (see also Ref. \cite{Fernandes2023}). In this case, the order parameter is governed by the action 
\begin{eqnarray}
S_\phi&=&\frac{1}{2}\sum_{i=1,2}\int_x \phi_i(x)\left(r_0-c^{-2}\partial_\tau^2-\nabla^2\right)\phi_i(x) \nonumber \\
&+&u\int_x \left(\phi_1(x)^2+\phi_2(x)^2\right)^2\nonumber \\
&+&v\int_x \left(\phi_1(x)^2+\phi_2(x)^2\right)^3 \nonumber \\
&+&w\int_x \phi_1(x)^2 \left(\phi_1(x)^2-3\phi_2(x)^2\right)^2.
    \label{eq:Sphi2}
\end{eqnarray}
This corresponds to the six-state clock model, and as such the ground state is six-fold degenerate with the relative amplitude between $\phi_1$ and $\phi_2$ obtained by minimizing the last term in the action above. In the remainder of the paper, we will focus on the case of an Ising-like magnetic multipolar order parameter.

At long wavelengths, we can write the elastic action in terms of longitudinal and transverse phonon modes \cite{karahasanovic2016elastic} 
\begin{equation}
    \mathcal{S}_\varepsilon=-\frac{1}{2}\sum_{\nu=L,T} \int_x u_\nu(x)\left(\partial_\tau^2+v_\nu^2\nabla^2\right) u_\nu(x).
      \label{eq:Su}
\end{equation}
Here $\nu=L$ and $T$ corresponds to longitudinal and transverse phonons with displacement $u_L$ and $u_T$ and velocity $v_L$ and $v_T$, respectively. The velocities  depend upon the elastic constants of the system and, for the tetragonal crystal under consideration, on the polar angle $\theta$ of the momentum.
\begin{eqnarray}
     v_T^2& = &\frac{1}{4} (c_{11}-c_{12}+2c_{44}\nonumber\\
     & + & (-c_{11}+c_{12}+2c_{44})\cos\left(2 \theta\right)),\label{eq::vT}\\
     v_L^2 & = & \frac{1}{2} (c_{11}+c_{44}+(-c_{11}+c_{44})\cos\left(2\theta\right)).\nonumber
\end{eqnarray}
For the system to be stable, it must follow that $v_L> v_T$. Note that, for three-dimensional crystals, there is an additional transverse mode. Since we are interested in the  point group operations relevant for the $B_{1g}^{-}$ order parameter $\phi$, the crucial lattice displacements are in-plane. Therefore, hereafter we focus on a single transverse mode with predominant in-plane polarization. 

Finally, using the fact that $\varepsilon_{ij}=(\partial_i u_j+\partial_j u_i)/2$, we can rewrite the dynamic coupling in Eq.\eqref{eq:altmelast} in terms of the longitudinal and transverse displacements. After Fourier transformation to momentum and Matsubara frequencies, we find:
\begin{eqnarray}
    \mathcal{S}^{\rm dyn}_{\rm c}&=&\lambda_0\int_{q} \frac{\omega_n}{|\boldsymbol{q}|}\phi\left(q\right)\left[\left(q_x^2-q^2_y \right) u_L(-q) \right.
    \nonumber\\
& - &    \left. 2q_xq_y u_T(-q)\right].\label{eq::Couple}
\end{eqnarray}
Here $\omega_n=2\pi n T$ are Matsubara frequencies and $q=(\boldsymbol{q},\omega_n)$.    
The coupling is anisotropic, since at $\boldsymbol{q}=(q_x,0)$ and  $\boldsymbol{q}=(0,q_y)$ only the longitudinal phonon couples to the multipolar order parameter. On the other hand, for $\boldsymbol{q}=(q_x,\pm q_x)$, the coupling is solely to the transverse phonons. This symmetry-selective interaction can also be used  to determine the nature of the multipolar magnetism of unknown symmetry. If one studies the spectrum of a phonon  of well-defined symmetry and observes traces of a magnon mode that vanish along specific high symmetry directions, one can deduce the symmetry of the altermagnon.

\subsection{Spectral functions and hybridized collective modes}\label{sec::Spectral}
For $d=3$ and $T=0$, the quartic interaction $u$ in Eq. \eqref{eq:Sphi} is marginally irrelevant. Below we will account for the corresponding logarithmic divergencies using a renormalization group approach. Anticipating the result of this analysis, we account for the effects of the dynamic coupling on the collective elastic-altermagnetic modes by considering an effective Gaussian theory with $r_0$ replaced by the renormalized mass $r$. This allows us to analyze the spectral properties that follow from the coupling in Eq.~\eqref{eq:altmelast} or, equivalently, Eq.~\eqref{eq::Couple}.
This is accomplished by integrating out one set of degrees of freedom and then calculating the propagator for the remaining one. 

\begin{figure}
    \includegraphics[width=0.48\textwidth]{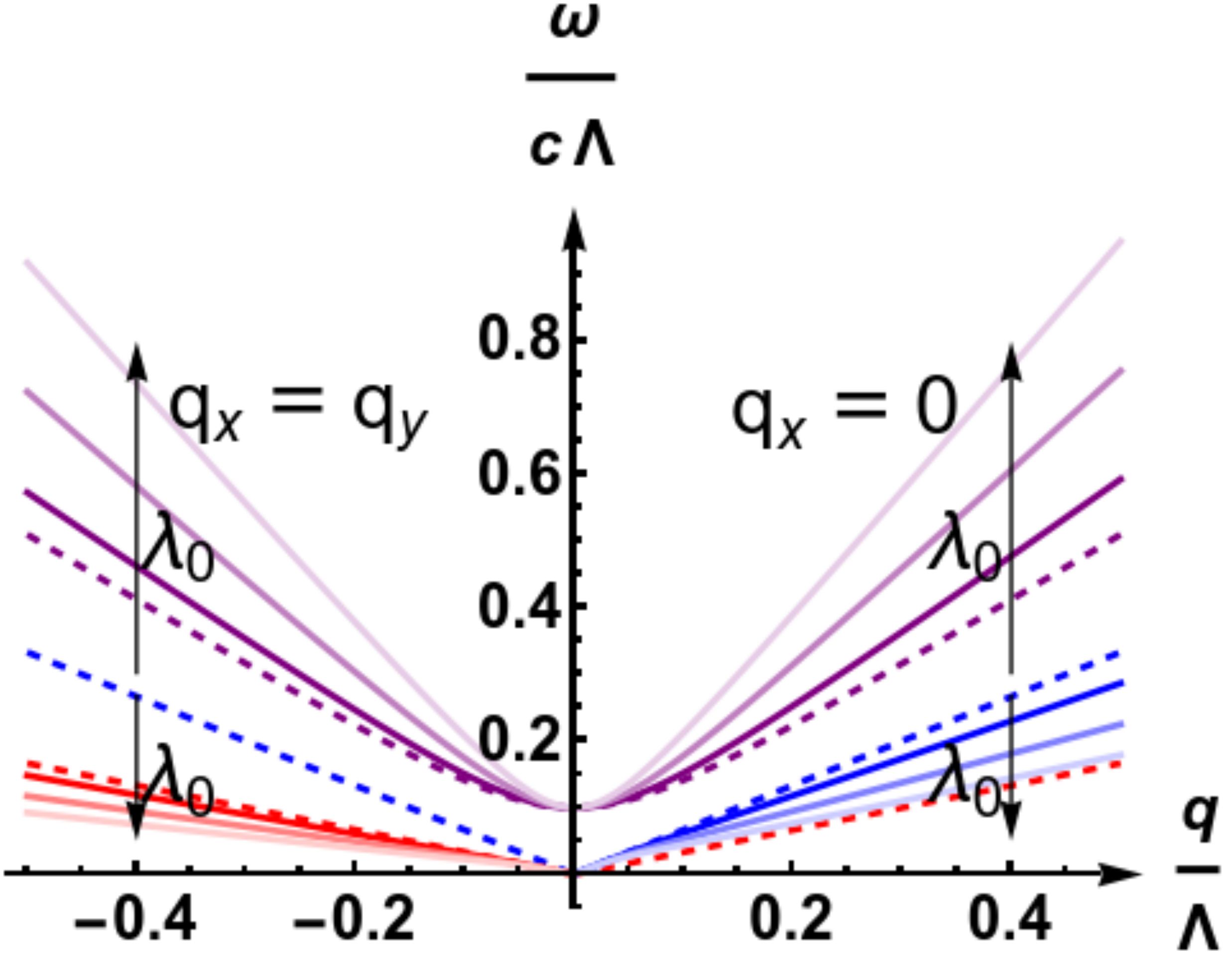}
    \caption{Altermagnon-like (purple), longitudinal  phonon-like (blue), and transverse phonon-like (red) dispersions for  momentum directions along high-symmetry lines in the $q_z=0$ planes, plotted for different values of the coupling strength $\lambda_0$. The dispersions of the phonon-like and altermagnon-like modes are strongly influenced by the coupling. The dashed lines show the ``pure'' modes without coupling. The modes mix to a dynamic altermagnon-polaron at large coupling values, leading to a softening of the phonon-like modes for given directions and a hardening of the altermagnon-like mode. The modes are shown for $\lambda_0=0,0.5,1,1.5$ with the shades getting progressively lighter with increased coupling constants. We set here $c/v_L=3/2$ and $c/v_T=3$. The altermagnon gap is given by $\sqrt{\frac{r}{\Lambda^2}}=0.1$; $\Lambda$ is a cutoff, as explained in the main text.}
    \label{fig::fulldisp}
\end{figure}

Starting with the altermagnetic propagator, we note that in the absence of the dynamic coupling, the collective modes in the disordered phase are gapped altermagnons (i.e. alter-paramagnons) with velocity $c$, as given by Eq. (\ref{eq:Sphi}). After integrating out the two phonon modes and performing the analytic continuation to real frequencies $i\omega_n \rightarrow \omega +i0^+$, we obtain the Gaussian renormalized altermagnon propagator: 
\begin{equation}
\chi(\boldsymbol{q},\omega)=\frac{1}{r+\boldsymbol{q}^2-\frac{\omega^2}{c^2}\Delta(\boldsymbol{q},\omega)},
\label{eq:aux_chi}\end{equation}
where
\begin{equation}
    \Delta(\boldsymbol{q},\omega)=1-\frac{c^2\lambda_0^2}{\boldsymbol{q}^2}\left(\frac{\left(q_x^2-q^2_y\right)^2}{\omega^{2}-v_L^2\boldsymbol{q}^2}+\frac{4q_x^2q^2_y}{\omega^{2}-v_T^2\boldsymbol{q}^2}\right).\label{eq::delta}
\end{equation}
The poles of this propagator in the case $c>v_{L,T}$ are shown in Fig.\ref{fig::fulldisp}. Where we considered $q_z=0$ such that $\theta=\frac{\pi}{2}$ and the phonon velocities are constant. This regime is relevant whenever electronic energy scales are larger than the lattice ones. For completeness, we discuss the opposite regime below. 

We see in Fig. \ref{fig::fulldisp}  that the modes which initially correspond to phonons also appear in the altermagnetic propagator. This is a direct consequence of the hybridization promoted by the dynamic coupling. As such, the phonon and alter-paramagnons are not independent modes, but hybridize into new collective modes that we dub paramagnon-polarons, inspired by the nomenclature used in Ref. \cite{flebus2017magnon}. Fig. \ref{fig::fulldisp} shows that, as the dynamic coupling increases, the phonon-like modes are softened whereas the gapped altermagnon-like mode is hardened. The coupling is clearly anisotropic, as only one phonon mode is softened along each of the high-symmetry directions considered in Fig. \ref{fig::fulldisp} -- namely, $q_x = 0$ (positive horizontal axes) and $q_x = q_y$ (negative horizontal axes). Away from the critical point ($r>0$), the altermagnon-like mode acquires a gap, whereas at the critical point, three gapless modes would emerge. 

\begin{figure}
     \centering
         \centering
         \includegraphics[width=0.48\textwidth]{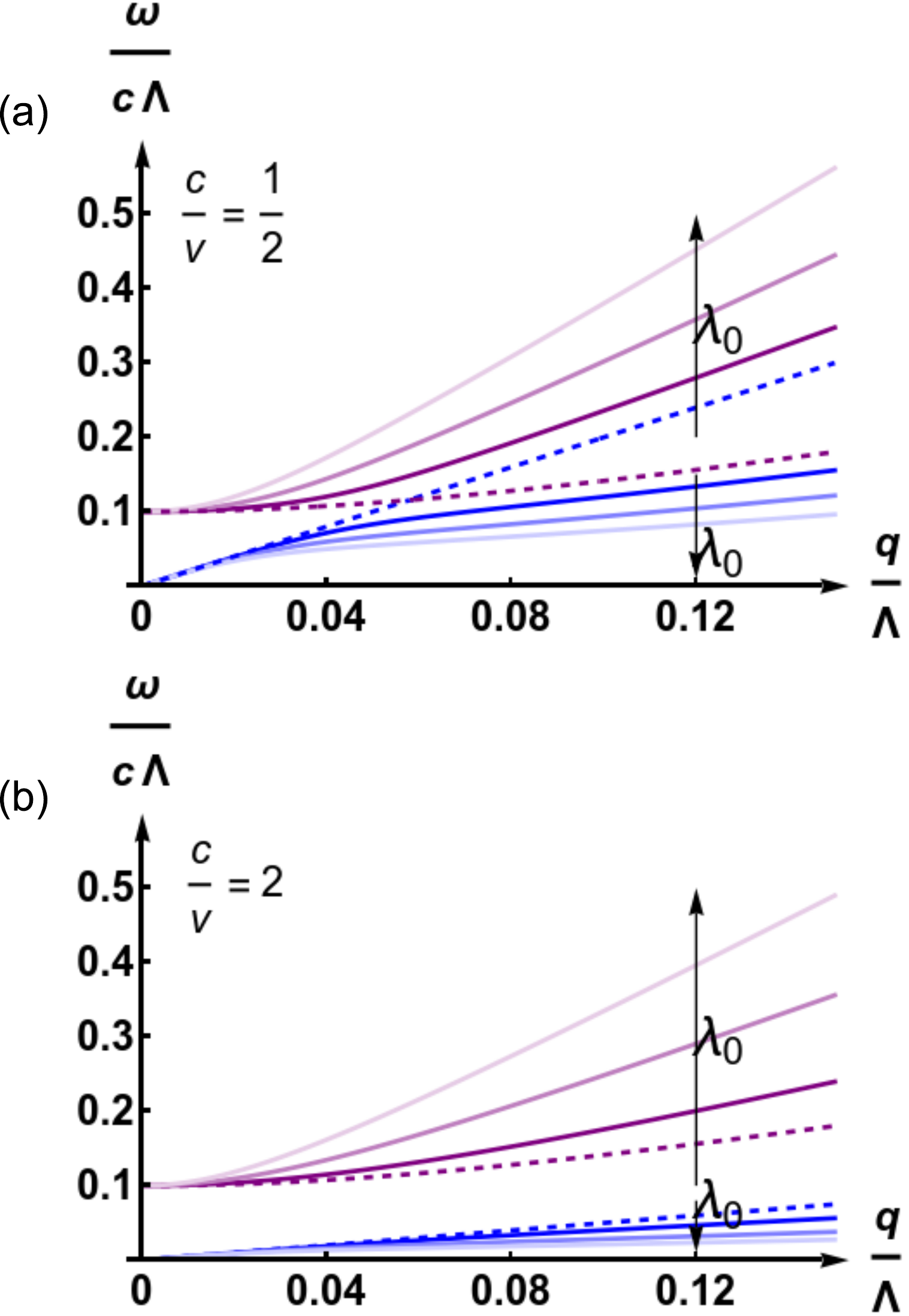}
        \caption{Dispersions of the altermagnon-like (purple) and longitudinal phonon-like (blue) mode for coupling constant values $\lambda=0,1,2,3$ along the $q_x=0$ direction. The dashed lines show the uncoupled modes and the solid lines show the hybridized modes. Panel (a) refers to the case $c/v_L=\tfrac{1}{2}$, whereas panel (b) refers to $c/v_L=2$. In panel (a), there is a level repulsion whereas in panel (b), we see the modes softening discussed in greater detail in Fig.\ref{fig::fulldisp}. In both panels, the altermagnon gap is given by $\sqrt{\frac{r}{\Lambda^2}}=0.1$.}
        \label{fig::levelrepulsion}
\end{figure}

We emphasize that this result is valid in the regime $c>v_{L/T}$. In the other regime $c<v_{L/T}$ we find instead a level repulsion and a greatly diminished softening of the phonon mode, as illustrated in Fig. \ref{fig::levelrepulsion}. Such a regime is applicable in materials with a very small magnetic interaction $J$. The regime $c>v_{L/T}$ is more common and forms the focus of this paper, although we will make reference to both regimes.

\begin{figure*}
     \centering
         \centering
         \includegraphics[width=0.75\textwidth]{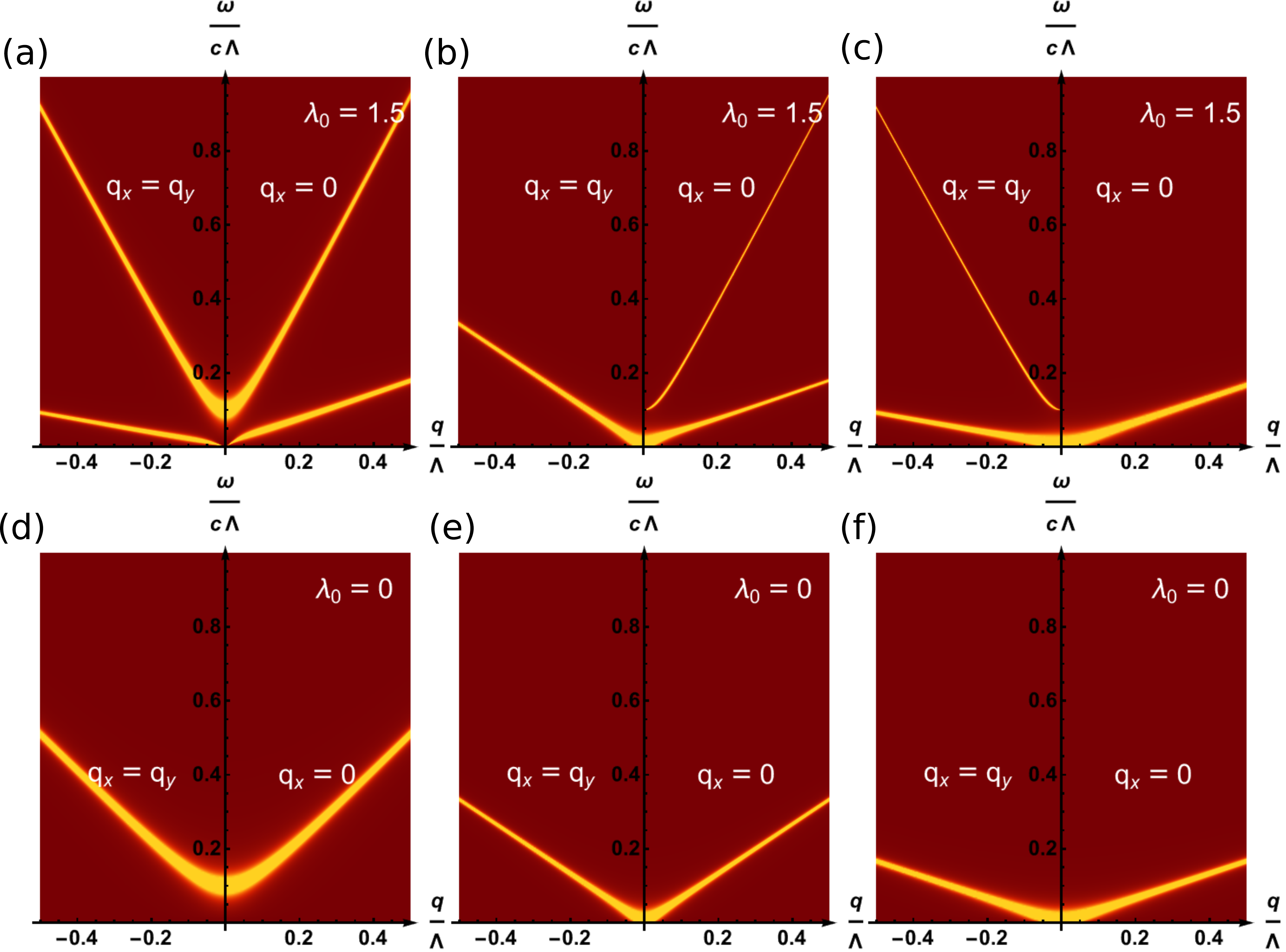}
        \caption{The spectral function for altermagnon (left panels), longitudinal phonon (middle panels) and transverse phonon (right panels) propagators as function of frequency $\omega$ and momentum $\mathbf{q}$ along two high-symmetry lines at the $q_z=0$ plane. Panels (a-c) refer to a non-zero coupling constant value  $\lambda_0= 1.5$, whereas panels (d-f) correspond to $\lambda_0=0$.
        When coupled (a-c), the modes hybridize and it is possible to detect the altermagnon by measuring the phonon spectrum, except along the special directions for which the phonon mode decouples from the altermagnon. This mixing is most prominent close to the critical point where the $\mathbf{q}=\mathbf{0}$ altermagnon excitation energy is small. We have set here $c/v_L=\frac{3}{2}$ and $c/v_T=3$. The altermagnon gap is given by $\sqrt{\frac{r}{\Lambda^2}}=0.1$. }
        \label{fig::spec}
\end{figure*}
After calculating all three propagators $\chi$, we can obtain the corresponding spectral functions by computing $\text{Im}\left(\chi\right)$ from the propagators:
\begin{equation}
\chi_T\left(\boldsymbol{q},\omega\right)=\frac{1}{\Delta_{T}\left(\boldsymbol{q},\omega\right)-\zeta_T\left(\boldsymbol{q},\omega\right) \omega^2+v_T^2\boldsymbol{q}^2},
\end{equation}
where $\chi_T$ is the propagator for the transverse phonon with renormalized coefficient 
\begin{equation}
\zeta_T\left(\boldsymbol{q},\omega\right)=1+\frac{\lambda_0^2}{\boldsymbol{q}^2}\frac{q_x^2q_y^2}{r_0+\boldsymbol{q}^2-\frac{\omega^2}{c^2}},
\end{equation}
of the dynamic term, while $\Delta_{T}\left(\boldsymbol{q},\omega\right)$ corresponds to coupling of the two phonons away from both high symmetry directions
\begin{equation}
\Delta_{T}\left(\boldsymbol{q},\omega\right)=\frac{\left(\frac{4\lambda_0^2\omega^2q_xq_y\left(q_x^2-q_y^2\right)}{\boldsymbol{q}^2\left(r_0+\boldsymbol{q}^2-\frac{\omega^2}{c^2}\right)}\right)^2}{-\zeta_L\left(\boldsymbol{q},\omega\right) \omega^2+v_L^2\boldsymbol{q}^2},
\end{equation}
with
\begin{equation}
   \zeta_L\left(\boldsymbol{q},\omega\right)=1+ \frac{\lambda_0^2}{\boldsymbol{q}^2}\frac{\left(q_x^2-q_y^2\right)^2}{r_0+\boldsymbol{q}^2-\frac{\omega^2}{c^2}}.
\end{equation}
For the longitudinal phonon propagator we have
\begin{equation}
\chi_L\left(\boldsymbol{q},\omega\right)=\frac{1}{\Delta_L\left(\boldsymbol{q},\omega\right)-\zeta_L\left(\boldsymbol{q},\omega\right)\omega^2+v_L^2\boldsymbol{q}^2},
\end{equation}
where
\begin{equation}
    \Delta_L\left(\boldsymbol{q},\omega\right)=\frac{\left(\frac{4\lambda_0^2\omega^2q_xq_y\left(q_x^2-q_y^2\right)}{\boldsymbol{q}^2\left(r_0+\boldsymbol{q}^2-\frac{\omega^2}{c^2}\right)}\right)^2}{-\zeta_T\left(\boldsymbol{q},\omega\right)\omega^2+v_T^2\boldsymbol{q}^2}.
\end{equation}
To model the finite lifetimes arising from damping or other processes, we add a small imaginary part to the frequency $\omega$ on the real axis. Fig. \ref{fig::spec} shows the spectral functions as a density plot. For the altermagnetic propagator we see that along each high-symmetry direction, only one phonon-like mode has a non-zero spectral weight, whereas the altermagnon-like mode is symmetric. This effect mirrors the behavior of the poles of the altermagnetic propagator discussed above in Fig. \ref{fig::fulldisp} and is a direct consequence of Eq.\ref{eq::Couple}. For $q_x = \pm q_y$, the altermagnon only couples to the longitudinal phonon, whereas for $q_x = 0$ or $q_y=0$, the altermagnon couples to the transverse mode, making this mode visible.  
The anisotropy of the dynamic coupling is also manifested in the gapped altermagnon-like mode when we plot the spectrum of the longitudinal phonon. Because this phonon mode does not hybridize with the altermagnon along $q_x=q_y$, it can only be observed in certain directions. Similarly, in the spectrum of the transverse mode, the gapped altermagnon-like mode has a vanishing spectral weight along the $q_x = 0$ direction. In either case, we also see that along the directions where the altermagnon-phonon coupling is non-zero, the gapless phonon-like mode softens.
These results show that even though directly measuring an altermagnon is a nontrivial task, by measuring the phonon spectrum at finite momentum, even away from the critical point, it is possible to assess the altermagnon mode -- provided the measurement is along a specific momentum-space direction. Conversely, by measuring the phonon spectrum along high-symmetry directions and identifying which ones display a gapped mode allows one to obtain the symmetry of the altermagnetic order parameter.
\begin{figure*}
     \centering
         \centering
         \includegraphics[width=\textwidth]{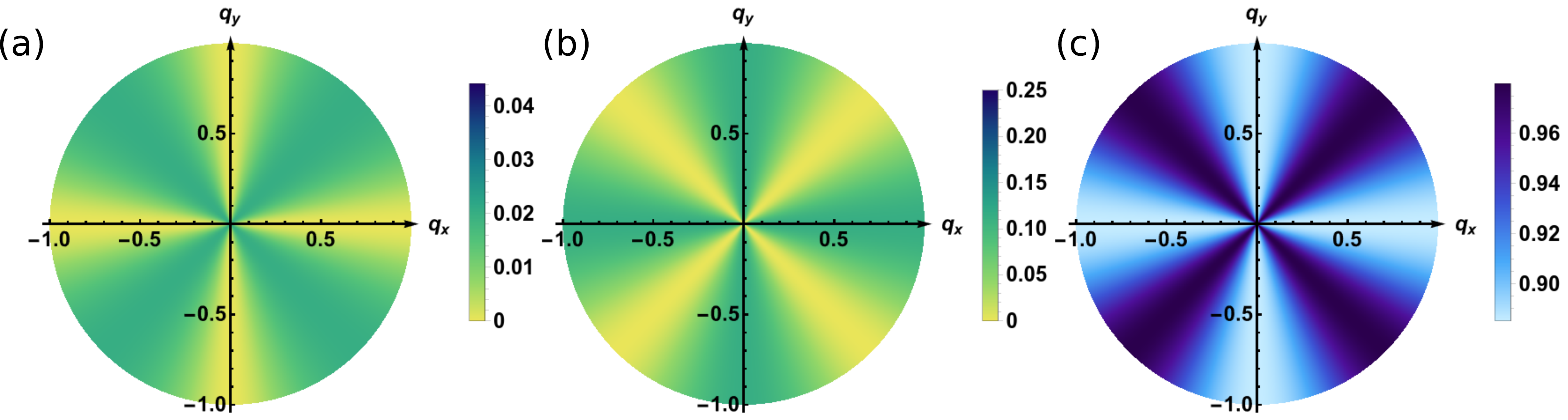}
        \caption{Anisotropic momentum dependence of the spectral weights of the transverse phonon-like (a), longitudinal phonon-like (b), and altermagnon-like (c) branches of the altermagnetic propagator for a non-zero coupling $\lambda_0=0.5$. The phonon-like modes decouple from the altermagnon-like one along high symmetry directions. These figures are plotted for $r=0$, $c/v_L=\frac{3}{2}$ and $c/v_T=3$}
        \label{fig::AiL}
\end{figure*}
In order to further illustrate the anisotropic nature of the coupling between the altermagnetic and phonon modes, we show a density plot of the spectral weight of each branch of the altermagnetic spectral function along the entire $q_x$, $q_y$ plane in Fig. \ref{fig::AiL}. We see that the weight of the longitudinal phonon vanishes along $q_x=\pm q_y$, as along these directions the phonon decouples and no longer contributes to the altermagnon propagator. Analogously, the weight of the transverse  phonon no longer contributes along $q_{x,y}=0$, such that the weight of this branch vanishes along these directions. While the spectral weight of the altermagnon branch of course never drops to zero, it becomes fourfold anisotropic.

In the regime where at least one of the phonon velocities is larger than the velocity of the altermagnon excitations the two modes cross with level repulsion for generic momenta as long as one is not right at the altermagnetic quantum critical point. However, for specific directions in momentum space the gap that opened vanishes. For the longitudinal phonons this is the case along the diagonal $q_x=\pm q_y$ and for the transverse modes along the axes, i.e. $q_x=0$ or $q_y=0$. Hence, nodal lines of the gap form.  The magnitude of the in-plane component of the momentum at this nodal line is $\left|\boldsymbol{q}_\parallel \right|\sim\frac{r^{1/2}}{\sqrt{v_{L,T}^{2}/c^{2}-1}}$. If one expands the dispersion near these crossing points one obtains a linear dispersion  similar to the one of a Weyl system with vanishing gap. These lines are protected by the crystalline symmetry of the system, but may nevertheless have implications for surface states of the combined spectrum. 

 The  coupling between the canonical momentum of one degree of freedom  and the coordinate of another one is at the heart of our discussion. As we considered effective Gaussian theories, interesting insight can be gained by considering a simple Hamiltonian that describes two coupled oscillators in which the displacement of one is coupled to the momentum of the other: 
\begin{equation}
    H=\sum_{i=1,2}\left(\frac{p_{i}^{2}}{2m_{i}}+\frac{m_{i}\omega_{i}^{2}}{2}x_{i}^{2}\right)+\frac{\lambda}{2}p_{1}x_{2}.
\end{equation}
As usual, the problem can be diagonalized using a $4\times 4$ symplectic matrix $\cal{S}$.
The transformation $\left(x_{1},x_{2},p_{1},p_{2}\right)\rightarrow{\cal S}^{-1}\left(x_{1},x_{2},p_{1},p_{2}\right)$
can also be cast as a unitary transformation of the operators, such
as $x_{i}\rightarrow Ux_{i}U^{-1}$ where
\begin{equation}
U=e^{i\left(ap_{1}p_{2}+bx_{1}x_{2}\right)},
\end{equation}
where the coefficients $a$ and $b$ can be expressed in terms of the $m_i$,$\omega_i$, and the coupling $\lambda$.  Applied to the vacuum, such an operator creates two-modes squeezed
states made by the two coupled oscillators \cite{hong2007one}. The two-mode squeezing occurs regardless of whether the magnetic system is gapped or not. It reflects the fact that the momentum-coordinate coupling strongly changes the relative fluctuations of the involved degrees of freedom, where the softening of one mode enforces the hardening of the other.

\section{Impact of the dynamic strain coupling on the altermagnetic phase diagram}\label{sec::phase}
 Having established how the collective altermagnetic and phonon modes are hybridized by the dynamic strain coupling, we now discuss how the altermagnetic phase transition is impacted by this coupling. In order to determine the phase diagram, we perform a one-loop renormalization group (RG) calculation for the quartic coefficient $u$ and the mass term coefficient $r$. We integrate out the phonon modes and obtain 
\begin{equation}
S=\frac{1}{2}\int_q \phi\left(q\right)\chi^{-1}(q)\phi\left(-q\right)+u\int\phi\left(x\right)^4,
\end{equation}
with the phonon-renormalized inverse altermagnon propagator on the real frequency axis written in Eq. \ref{eq:aux_chi}. The form of coupling Eq. \ref{eq::delta} implies rather different behaviors in the regimes where $\omega$ is small or large compared to $c q$. Considering first $\omega \gg v_{T,L}\left|\boldsymbol{q}\right|$, we have
\begin{equation}
    \Delta(\boldsymbol{q},\omega) \approx 1-c^2\lambda_0^2\frac{\boldsymbol{q}^2}{\omega^2},
\end{equation}
such that the coupling renormalizes the coefficient of the $q^2$ term of the altermagnetic propagator. Since the $q^2$ coefficient is proportional to the inverse squared correlation length, its suppression implies an enhancement of the spatial fluctuations mediated by the dynamic strain coupling. On the other hand, when $\omega\ll v_{L,T}\left| \boldsymbol{q}\right|$ we find
\begin{eqnarray}
    \Delta(\boldsymbol{q},\omega)&\approx & 1+\frac{c^2\lambda_0^2}{v_T^2},\qquad q_x=q_y ,  \\
       \Delta(\boldsymbol{q},\omega)&\approx & 1+\frac{c^2\lambda_0^2}{v_L^2},\qquad q_x=0 .
\end{eqnarray}
In this regime, it is the altermagnon velocity  $c$ that is renormalized downwards by the coupling, which suppresses quantum fluctuations.

As our goal is to calculate the phase diagram at nonzero temperatures, we employ the crossover method outlined in detail in Ref. \cite{d2003low}. We start from the flow equations given by a perturbative RG calculation, which is controlled at the upper critical dimension $d=3$ and includes the logarithmic corrections beyond mean field theory.  
\begin{eqnarray}
\frac{dr}{dl} & = & 2r+3u\frac{d}{dl}\int_{q}^{>}\chi_{0}\left(q\right)-3ur\frac{d}{dl}\int_{q}^{>}\chi_{0}^{2}\left(q\right),\nonumber \\
\frac{du}{dl} & = & -9u^{2}\frac{d}{dl}\int_{q}^{>}\chi_{0}^{2}\left(q\right).
\end{eqnarray}
where $\chi_0$ is the propagator (Eq. \ref{eq:aux_chi}) taken at $r=0$. To derive these expressions, we have integrated out the long wavelength modes leaving us with a set of shell integrals over $\Lambda e^{-l}<\left|\boldsymbol{q}\right|<\Lambda$ for some momentum cutoff $\Lambda$, with $l$ a parameter used to vary length-scales. Note that we employ no cutoff for the Matsubara frequencies. The shell equations can be solved in a straightforward way:
\begin{eqnarray}
    \begin{split}
\frac{d}{dl}\int_{q}^{>}&\chi_0^m\left(q\right) = \frac{d}{dl}\int_{\Lambda e^{-l}<\left|\boldsymbol{q}\right|<\Lambda}\frac{d^{3}q}{\left(2\pi\right)^{3}}T\sum_{n=-\infty}^{\infty}\chi_0^m\left(q\right) \\
& = \frac{\Lambda^{3}}{\left(2\pi\right)^{3}}T\sum_{n=-\infty}^{\infty}\int_{0}^{2\pi}\int^{\pi}_0 d\theta d\phi\sin\left(\theta\right) \chi_0^m\left(\Lambda\right),
    \end{split}
\end{eqnarray}
where we suppress the dependency of $\chi_0$ on $\omega_n$, $\phi$ and $\theta$. The flow equations at one-loop level are then
\begin{eqnarray}
\frac{dm}{dl} & = & 2m+g\Lambda cF_1\left(T\right)-gmc^3\Lambda^3 F_2\left(T\right),\nonumber \\
\frac{dg}{dl} & = &-3g^2\Lambda^3c^3F_2\left(T\right), \nonumber \\
\frac{dT}{dl} & = & T .
\end{eqnarray}
where we defined the dimensionless quantities
\begin{eqnarray}
  m=\frac{r}{\Lambda^2},\quad g=\frac{3uc\Lambda^{d-3}}{\left(2\pi\right)^d} ,
\end{eqnarray}
and
\begin{equation}
     F_m\left(T \right) =T\sum_{n=-\infty}^{\infty}\int_{0}^{2\pi}\int^{\pi}_0 d\theta d\phi\sin\left(\theta\right) \chi_0^m\left(\Lambda,\omega_n,\theta,\phi\right).
\end{equation}
Note that $T$ is the running temperature, whereas the physical temperature corresponds to $T(l=0)$.

Taking the limit $T\rightarrow 0$ and solving the flow equations naturally yields a phase diagram with a quantum critical point. 
We can define the distance from the critical point \cite{rudnick1976equations}
\begin{equation}
     t=m+\frac{c\Lambda}{2}F_1\left(0\right)g.
\end{equation}

\begin{figure}
     \centering
         \includegraphics[width=0.45\textwidth]{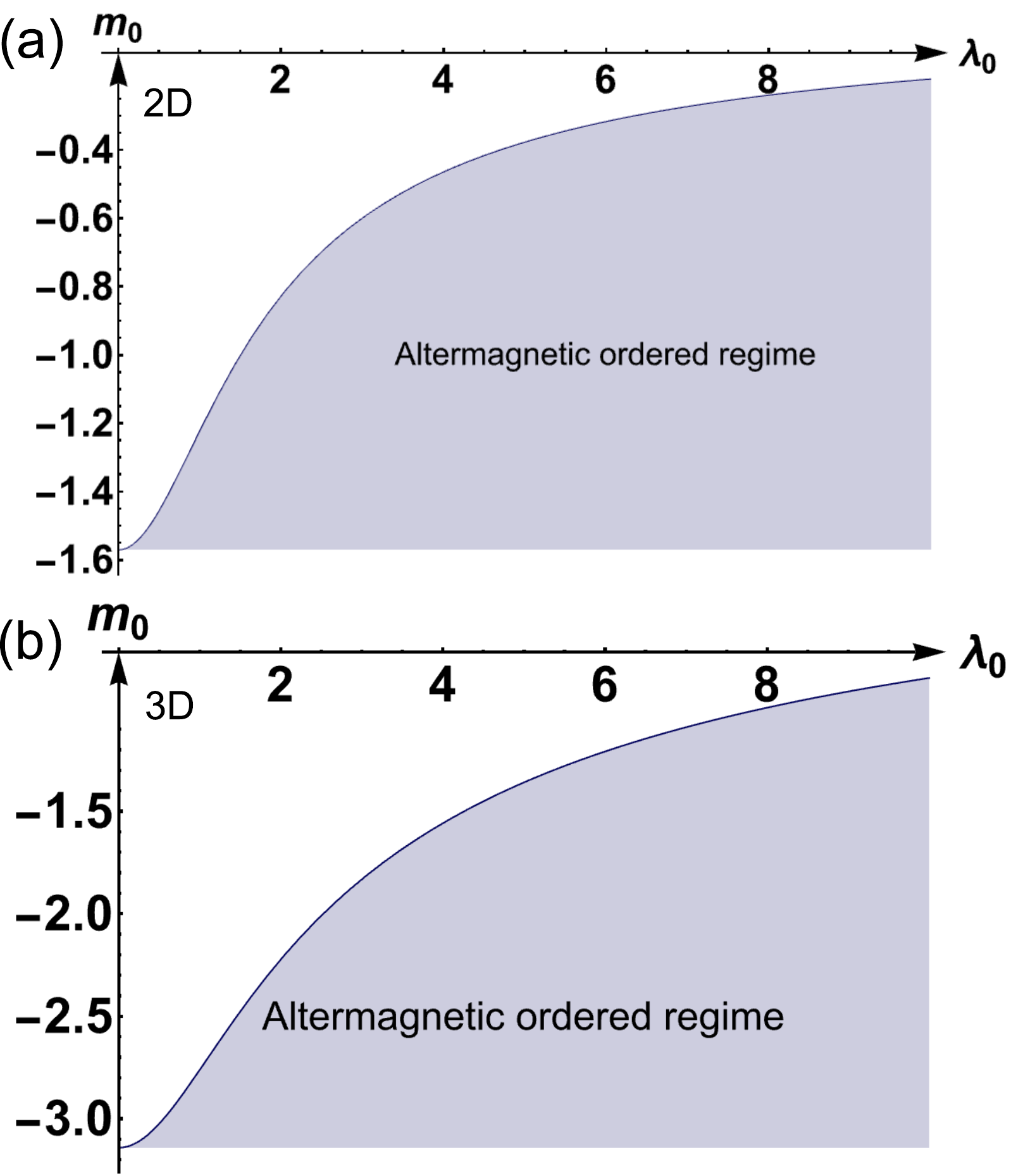}
         \caption{Ground-state (i.e. $T=0$) phase diagram of altermagnetic order as a function of the control parameter $m_0=r_0/\Lambda^2$ and the coupling constant $\lambda_0$ with $c>v_{L,T}$, Eq. (\ref{eq:v_smaller}) for the cases (a) 2D and (b) 3D.
         We see that at zero temperature, the formation of a dynamic altermagnon-polaron expands the ordered state (shaded region), since the altermagnetic transition occurs for higher values of $m_0$.
         }
         \label{fig::phaseR}
\end{figure}

We plot the phase diagrams at $T=0$ in Fig. \ref{fig::phaseR} for the two regimes. For the case $c > v_{L,T}$; we used the parametrization:
\begin{align}
\frac{v_T(\theta)}{c} &= \frac{1}{6}\sqrt{\frac{9}{2}+\frac{7}{2}\cos 2\theta} ,\nonumber \\
\frac{v_L(\theta)}{c} &= \frac{1}{3}\sqrt{\frac{3}{2}+\frac{1}{2}\cos 2\theta}.
\label{eq:v_smaller}
\end{align}

As for the case $c<v_{L,T}$, we considered the following parametrization:
\begin{align}
\frac{v_T(\theta)}{c} &= \frac{1}{6}\sqrt{450 + 350\cos 2\theta}, \nonumber \\
\frac{v_L(\theta)}{c} &= \frac{1}{3}\sqrt{150+50\cos 2\theta}.
\label{eq:v_larger}
\end{align}

The key result is that increasing the dynamic strain coupling constant $\lambda_0$ expands the regime with long-range altermagnetic order. This is consistent with what we found above that, in the quantum regime, $\lambda_0$ renormalizes $c$ and suppresses quantum fluctuations. Thus, larger values of the coupling leads to a larger ordered regime characterised by a smaller $|m_0|$ due to the hardening of the hybridized altermagnon-like mode. Hence, the dynamic coupling to phonons suppresses quantum altermagnetic fluctuations, reinforcing altermagnetic order. This is the case for both regimes $c<v_{L,T}$ and $c>v_{L,T}$.

We can also solve the flow equations for the case of small but finite $T$, which yields an expression for the transition temperature $T_c$. This crossover regime is relevant when $g_0=g(l=0)\ll 1$ and $T\ll c\Lambda$. In this regime, to first-order in $g_0$, it is sufficient to simply use the $T=0$ solution for $g\left(l\right) $ \cite{d2003low}
\begin{equation}
g\left(l\right)=\frac{g_0}{1+3c^3\Lambda^3F_2\left(0\right)g_0l}.
\end{equation}
Thermal fluctuations only enter at $\mathcal{O}\left(g_0^2\right)$. We now turn our attention to $m$, and consider the ansatz
\begin{equation}
m=m_0e^{\xi\left(l\right)}h\left(l\right).
\end{equation}
This ansatz solves our flow equation when
\begin{equation}
\begin{split}
\xi\left(l\right)&=2l-\Lambda^3c^3\int_0^lg\left(l'\right)F_2\left(l'\right)dl',\\
h\left(l\right)&=1+\frac{\Lambda c}{m_0}\int_0^le^{-\xi\left(l'\right)}F_1\left(l'\right)g\left(l'\right)dl'.
\end{split}
\end{equation}
We can then substitute these expressions in the flow equation for $m$ and integrate by parts to find an expression for the transition temperature $T_c$ to first-order in $g_0$. At the critical point, by definition, We can also set $l\rightarrow\infty$. We find:
\begin{equation}
\begin{split}
&m^c_0+\frac{\Lambda cg_0}{2}\int_0^{2\pi}\int_0^{\pi}d\theta d\phi \sin\left(\theta\right)\sum_{i=1}^3\frac{A_i}{2E_i}\\
&+g_0\Lambda cT_c^2\int_{0}^{\pi}\int_0^{2\pi}d\phi d\theta\sin\left(\theta\right)\sum_{i=1}^3A_i\left(\frac{\pi^2}{6E_i^3}\right.\\
&\left.+\frac{1}{E_i^2T_c}\log
\left(1-e^{-\frac{E_i}{T_c}}\right)-\frac{1}{E_i^3}{\rm Li}_2\left(e^{-\frac{E_i}{T_c}}\right)\right)=0.
\end{split}
\end{equation}
where $E_i$ is the energy for each mode and $A_i$ the corresponding weight for the three branches, determined via 
\begin{equation}
    \chi(\boldsymbol{q},\omega)=\sum_i\tfrac{c^2A_i(\boldsymbol{q})}{\omega^2-E_i(\boldsymbol{q})^2}
\end{equation}
and ${\rm Li}_2\left(z\right)$ is the poly-logarithm.

As well as in 3D, there also exists 2D altermagnetic candidates \cite{Smejkal2022} such as the quasi-2D oxide insulator $\text{V}_2\text{Se}_2\text{O}$ \cite{ma2021multifunctional} and semimetal $\text{Cr}_2\text{O}$ \cite{chen2021room}. We can carry out this analysis in 2D using an epsilon expansion with $\epsilon=3-d=1$. The calculation then becomes very similar compared to the 3D case taken at $\theta=\frac{\pi}{2}$ and with the only angular integration being over $\phi$.

\begin{figure}
     \centering
         \centering
         \includegraphics[width=0.48\textwidth]{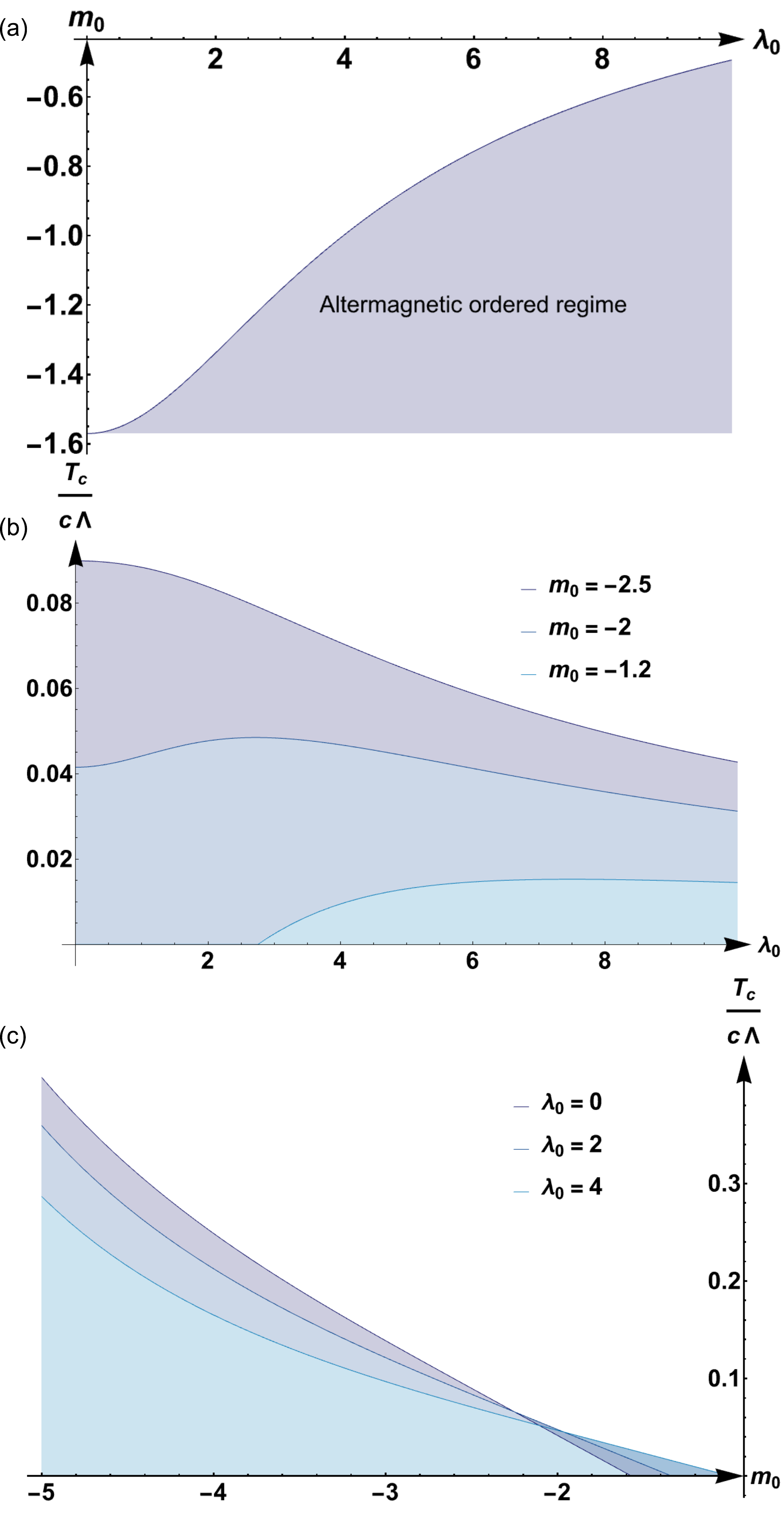}
         \label{fig::highVphase}
        \caption{(a) Ground state phase diagram for a 2D system. (b) Altermagnetic transition temperature $T_c$ as a function of the dynamic strain coupling constant $\lambda_0$ of Eq.~\eqref{eq:dyncoupl_intro} for different values of the bare altermagnon mass $m_0=r_0/\Lambda^2$. (c) Altermagnetic $T_c$ as a function of $m_0$ for different values of $\lambda_0$. These plots correspond to the case of a 2D system with $c<v_{L,T}$, Eq. (\ref{eq:v_larger}). Compared with Fig.\ref{fig::Phase}, which refers to the case $c>v_{L,T}$, the dynamic coupling to phonons has a weakened impact in the thermal regime, but still a significant impact in the quantum regime. The former is the result of a lack of softened phonons, whereas the latter is a consequence of the hardened altermagnon.
        }
        \label{fig::Phasev}
\end{figure}

Since this equation cannot be solved analytically for $T_c$, we resort to numerical methods to find the solution. We plot the obtained phase diagram in Fig. \ref{fig::Phase}, using the parametrization for the velocities of Eq. (\ref{eq:v_larger}), i.e. $c > v_{L,T}$. We see in 2D that, in general, increasing the dynamic coupling to phonons $\lambda_0$ leads to a decrease in the transition temperature. This is consistent with the results of Fig.\ref{fig::fulldisp}, which shows that the coupling leads to a softening of phonons. A consequence of this softening is that for larger coupling, the system contains a larger population of soft phonons, which suppresses altermagnetic order. As $T=0$ is approached this effect is less relevant, as there are no phonon modes occupied at zero temperature. By comparing Fig. \ref{fig::phaseR} and Fig. \ref{fig::Phase}, we note that, for a given coupling constant value, unless $m_0$ is within the $T=0$ ordered regime, there is no transition at non-zero temperature. Increasing $\lambda_0$ leads to an increase in $m_0^c$ and for any $m_0<m_0^c\left(\lambda_0\right)$, the transition temperature rises to a maximum before being suppressed. This initial rise is most likely due to the system being in the regime where the altermagnon hardening is still the dominant effect. While these conclusions refer to the case where $c>v_{L/T}$, we can also calculate the phase diagram for the other regime, $c<v_{L/T}$. Such a regime is less common but would be the case for systems with small magnetic interaction $J$. As discussed above in Fig. \ref{fig::levelrepulsion}, the impact of the dynamic coupling on the phonon-like mode is significantly diminished, whereas the altermagnon-like mode still hardens. Consequently, as shown in Fig. \ref{fig::Phasev}, where the parametrization of Eq. (\ref{eq:v_larger}) was used, in the regime $c<v_{L/T}$ we still find a very similar phase diagram near the quantum critical regime; however, in the thermal regime the coupling has a much smaller effect on the transition temperature. In 3D, the behaviour at the QCP remains the same but thermal fluctuations are less relevant, as we see in Fig.\ref{fig::Phase}. In this case, the suppression of quantum fluctuations is the dominant effect, such that an increase in coupling leads to a slow increase of the transition temperature. If the temperature is  high ($\frac{T_c}{\Lambda c}>>1$), one would potentially expect soft phonons to be relevant, however we find instead that the transition temperature starts to plateau; at high temperatures we are in the classical regime which is equivalent to taking only the zeroth Matsubara frequency and as such the coupling to phonons is zero. At high temperatures, the coupling hence has little to no effect on $T_c$.

\section{conclusions}
\label{sec:conclusions} 
In summary, we showed from symmetry considerations that a dynamic coupling between strain and the momentum of a magnetic collective mode naturally emerges in a class of systems with multipolar magnetic order. An important application of these results is for the case of altermagnets, as they are described by $d$-wave, $g$-wave, and $i$-wave magnetization order parameters, which in turn correspond to non-zero magnetic multipoles. While in this paper we focused on a tetragonal crystal with $D_{4h}$ symmetry and an altermagnetic order parameter transforming as the irreducible representation $B_{1g}^-$ (relevant for instance for the altermagnet candidate MnF$_2$), the results are more general, as we pointed out by commenting on crystals with $O_h$ point group. One of our main results is the demonstration that, due to this dynamic strain coupling, altermagnons can in principle be probed directly from the phonon spectrum. This is important, as detecting such a state with zero net-magnetisation via the magnetic spectrum is a challenging task.

The coupling discussed here can be understood as an internal, fluctuation-induced non-dissipative    response, which gives rise to stress $\sigma_{ij}$ generated  by a time-varying strain in the presence of the magnetic  multipolar collective mode.  It is analogous to the stress that occurs due to a finite Hall viscosity.
The coupling induces a symmetry-sensitive dynamic hybridization of phonon and altermagnon modes, i.e. an altermagnon-polaron. It softens the former and hardens the latter, giving rise to significant changes of regions where altermagnetism occurs in the temperature-quantum fluctation phase diagram. 
In both 3D and 2D systems at $T=0$, the effect of the coupling leads to an enhancement of order, hence in the $T=0$ plane of the phase diagram, the ordered regime is enlarged. 
At non-zero $T$ in a 2D system the situation changes. Now, thermal fluctuations (phonons) become the dominant effect, and the renormalization of these fluctuations due to the dynamic coupling leads to a high population of soft phonons. These, in turn, suppress order, leading to a reduction in the transition temperature.  In a 3D system, the ordered regime is also increased at finite-T as thermal fluctuations remain small. The results for a 2D system suggest that thermal fluctuations become the dominant effect at finite $T$ in highly anisotropic 3D systems.

While the focus of this paper was on altermagnets, it is important to note that the coupling in Eq. (\ref{eq:dyncoupl_intro}) should also be relevant for certain ferromagnets. The condition for this coupling to be present is that the magnetization and some of the strain components must transform as the same irreducible representation $\Gamma$ of the point group, the difference being that the former is time-reversal-odd ($\Gamma^{-}$) and the latter, time-reversal-even ($\Gamma^+$). While this is not possible in the cubic group $O_h$, it is allowed for tetragonal $D_{4h}$ and hexagonal $D_{6h}$ ferromagnets with in-plane moments. In those cases, respectively, the two-component in-plane magnetization transforms as $E_{g}^-$ and $E_{1g}^-$, whereas the out-of-plane shear strain doublet $\left(\varepsilon_{xz}, \varepsilon_{yz} \right)$ transforms as $E_{g}^+$ and $E_{1g}^+$. An even more promising class of systems is that of $D_{2h}$ orthorhombic ferromagnets. In these cases, each of the three components of the magnetization transform separately as one of the one-dimensional irreducible representations $B_{ig}^-$ with $i=1,2,3$. But it turns out that each of the three shear strains, $\varepsilon_{xy}$, $\varepsilon_{xz}$, and $\varepsilon_{yz}$, transforms as one of the $B_{ig}^+$ irreps. The same conclusions hold for the other two orthorhombic point groups, $D_2$ and $C_{2v}$. Therefore, the effects discussed here should be present in any orthorhombic ferromagnet. A promising family of materials to search for this effect are the ferromagnetic Mott insulating perovskites $A$TiO$_3$, with appropriate rare-earth $A$ \cite{Imada2004}.

\section{Acknowledgements}
We are grateful to B. Flebus, I. I. Mazin, J. Sinova, L. \v{S}mejkal, R. Valent\'{i} for helpful discussions. This work was supported by the Deutsche Forschungsgemeinschaft (DFG, German Research Foundation) through TRR 288, 422213477 Elasto-Q-Mat through projects A07 (J.S) and the DFG project SCHM 1031/12-1 (C.R.W.S.) R.M.F was supported by the Air Force Office of Scientific Research under Award No. FA9550-21-1-0423. R.M.F. also acknowledges a Mercator Fellowship from the German Research Foundation (DFG) through TRR 288, 422213477 Elasto-Q-Mat.

\appendix
\section{Alternative derivation of the dynamic coupling}\label{sec::coupling}
The coupling
of Eq. \ref{eq:altmelast} can also be obtained by following the approach
of Refs. \cite{Link2018,Rao2020}. Consider a fermionic field operator
$c\left(\boldsymbol{x}\right)$ that is a spinor in spin and orbital
space.  Performing a deformation of the lattice with non-symmetrized
strain $\widetilde{\varepsilon}_{\alpha\beta}=\partial_{\alpha}u_{\beta}$,
we consider a coordinate transformation
\begin{equation}
\boldsymbol{x'}=\hat{\Gamma}\left(t\right)^T\boldsymbol{x}
\end{equation}
where
\begin{equation}
\hat{\Gamma}\left(t\right)=e^{\widetilde{\varepsilon}}
\end{equation}
We consider an arbitrary strain field such that $\hat{\Lambda}\left(t\right)$ is an arbitrary matrix with a positive determinant. The fermionic field transforms as \cite{Link2018}
\begin{equation}
c_{\widetilde{\varepsilon}}\left(\boldsymbol{x}\right)=\sqrt{\det\Gamma}\,c\left(\Gamma^T\boldsymbol{x}\right)
\end{equation}
In the case of a small change, we have $\widetilde{\varepsilon}\rightarrow \widetilde{\varepsilon}+\delta \widetilde{\varepsilon}$
\begin{equation}
\frac{\partial}{\partial\widetilde{\varepsilon}_{\alpha\beta}}c_{\widetilde{\varepsilon}}\left(\boldsymbol{x}\right)=\frac{\delta_{\alpha\beta}}{2}c_{\widetilde{\varepsilon}}\left(\boldsymbol{x}\right)+x_{\alpha}'\frac{\partial}{\partial x'_{\beta}}c_{\widetilde{\varepsilon}}\left(\boldsymbol{x}\right)
\end{equation}
In order to study an infinitesimal change we set $\widetilde{\varepsilon}=0$ such that $\boldsymbol{x}'=\boldsymbol{x}$. In this case the infinitesimal transformation can be written as:
\begin{equation}
\begin{split}
c_{\widetilde{\varepsilon}}\left(\boldsymbol{x}\right)=\left[1-i\sum_{\alpha\beta}\widetilde{\varepsilon}_{\alpha\beta}\mathcal{L}_{\alpha\beta}\right]c\left(\boldsymbol{x}\right)
\end{split}
\end{equation}
where
\begin{equation}
\mathcal{L}_{\alpha\beta}=\frac{i\delta_{\alpha\beta}}{2}+ix_{\alpha}\frac{\partial}{\partial x_{\beta}}=-\frac{1}{2}\left(x_{\alpha}p_{\beta}+p_{\beta}x_{\alpha}\right).
\end{equation}
One can also include rotations that act in the internal space. We refer to the generator for this transformation as $\mathcal{S_{\alpha\beta}}$ and consider $\mathcal{J_{\alpha\beta}}=\mathcal{L_{\alpha\beta}}+\mathcal{S_{\alpha\beta}}$. For specific examples, see \cite{Link2018,Rao2020}.
The important point is that $\mathcal{J_{\alpha\beta}}$ is even under parity, odd under time reversal and its symmetric part transforms like a symmetric second rank tensor, i.e. just like a multipolar order parameter discussed in this paper.
The field operator $c_{\widetilde{\epsilon}}\left(\boldsymbol{r}\right)$
of the strained system is hence related to the unstrained case
via:
\begin{equation}
c_{\widetilde{\varepsilon}}\left(\boldsymbol{x}\right)=e^{-i{\rm Tr}\left(\widetilde{\varepsilon}^{T}{\cal J}\right)}c\left(\boldsymbol{x}\right)=U\left(t\right)c\left(\boldsymbol{x}\right),
\end{equation}
with strain generators ${\cal J}_{\alpha\beta}=-\frac{1}{2}\left(x_{\alpha}p_{\beta}+p_{\beta}x_{\alpha}\right)+\frac{i}{8}\left[\sigma_{\alpha},\sigma_{\beta}\right]$.
$\boldsymbol{x}$ and $\boldsymbol{p}=-i\nabla$ are the position
and momentum operators and the Pauli matrices $\sigma_{\alpha}$ act
in orbital space. 

Because $U\left(t\right)$ represents a time-dependent transformation, a term can be introduced into the action via
\begin{equation}
S_{{\rm c}}=-i\int d\tau d^{3}\boldsymbol{x} c^{\dagger}(\boldsymbol{x})U\left(t\right)\frac{d}{dt}\left(U\left(t\right)^{-1}\right)c(\boldsymbol{x}).
\end{equation}
The coupling term that emerges from the relation between the
strained and unstrained system is hence \cite{Link2018,Rao2020} $S_{{\rm c}}= \int d\tau d^{3}\boldsymbol{x} \sum_{\alpha\beta}\widetilde{\varepsilon}_{\alpha\beta}\frac{d}{d\tau}c^{\dagger}{\cal J}_{\alpha\beta}c$.
This is the coupling of strain to the time derivative of a fermionic bilinear that transforms like the multipolar magnetic order parameter. It is the analog of the coupling Eq. \ref{eq:altmelast} to the conjugate momentum that appears on the level of the Hamiltonian.

\bibliography{mybib}

\end{document}